\documentclass{article}
\usepackage{spconf,amsmath,graphicx}
\usepackage{amsmath,epsfig}
\usepackage{amsfonts,amssymb}

\newcommand{\bqn}{\begin{eqnarray}}
	\newcommand{\eqn}{\end{eqnarray}}
\newcommand{\bq}{\begin{eqnarray*}}
	\newcommand{\eq}{\end{eqnarray*}}

\newcommand{\ba}{\left( \begin{array}}
\newcommand{\ea}{\end{array} \right)}

\usepackage{url} 
\usepackage{xcolor}
\newcommand{\blue}[1]{{\color{blue} #1}}
\newcommand{\green}[1]{{\color{green} #1}}
\newcommand{\red}[1]{{\color{red} #1}}

\title{Counterfactual Analysis of Brain Network Dynamics
\thanks{Published in the IEEE International Symposium on Biomedical Imaging (ISBI), 2026.}}
\name{
Moo K. Chung$^1$
\quad Luigi Maccotta$^2$
\quad Aaron Struck$^2$
}

\address{
$^1$University of Wisconsin--Madison, USA\\
$^2$Washington University in St. Louis, USA
}

\begin{document}
\maketitle

\begin{abstract}
Causal inference in brain networks has traditionally relied on regression-based models such as Granger causality, structural equation modeling, and dynamic causal modeling.  
While effective for identifying directed associations, these methods remain descriptive and acyclic, leaving open the fundamental question of intervention: {\it what would the causal organization become if a pathway were disrupted or externally modulated?}  We introduce a unified framework for \emph{counterfactual causal analysis} that models both pathological disruptions and therapeutic interventions as an {\it energy-perturbation problem} on network flows.  Grounded in Hodge theory, directed communication is decomposed into dissipative and persistent (harmonic) components, enabling systematic analysis of how causal organization reconfigures under hypothetical perturbations.  This formulation provides a principled foundation for quantifying network resilience, compensation, and control in complex brain systems.
\end{abstract}

 \section{Introduction}
Understanding causality in brain networks is one of the central challenges in computational neuroimaging. Traditional approaches—including Granger causality, structural equation modeling (SEM), and dynamic causal modeling (DCM)—have provided powerful tools for testing directed influences between brain regions \cite{friston.2014}. 
 However, these methods remain fundamentally associational: they infer directed dependencies from observational data but do not provide a principled way to reason about {\it interventions}. Consequently,  the mechanistic question—{\it how would the brain’s causal architecture change if a specific pathway or feedback loop were disrupted?}—remains unresolved.

Counterfactual analysis provides a principled framework for answering ``what if'' questions that go beyond assocations to model the consequences of hypothetical interventions.  
It has emerged as a central theme across modern data science, from epidemiology \cite{vanderweele.2020} to machine learning \cite{dhinagar.2024}, where interventions are often infeasible or unethical to observe directly.  In neuroscience, counterfactual reasoning is increasingly recognized as critical for bridging observational neuroimaging data with mechanistic insight \cite{matsui.2022,dhinagar.2024}.  By explicitly simulating interventions—even when no experimental manipulation is possible—
 counterfactual analysis moves beyond descriptive connectivity toward predictive models of network stability, compensation, and control.

In this study, we develop a formal framework for counterfactual causal analysis grounded in \emph{Hodge theory} \cite{anand.2023.TMI} and the \emph{minimum-energy principle}.  
We represent directed functional interactions as energy-carrying edge flows and model perturbations as controlled changes in the network’s energy landscape.  
This allows us to test how causal organization would reconfigure following structural damage, neuromodulation, or adaptive reorganization—even when such interventions cannot be performed in vivo.  We demonstrate this framework through applications to temporal lobe epilepsy (TLE), comparing pathological recurrence with therapeutic disconnection.

\section{Methods}

\subsection{Data and preprocessing}

We analyzed resting-state fMRI (rs-fMRI) data from 400 participants of the Human Connectome Project (HCP) \cite{vanessen.2012}. Each rs-fMRI dataset was acquired with 2\,mm isotropic spatial resolution and consisted of 1200 time points. Preprocessing followed the HCP minimal preprocessing pipelines \cite{glasser.2016}, including motion correction, spatial normalization, and artifact removal. Time frames with framewise displacement exceeding 0.5\,mm, along with their adjacent volumes, were excluded from further analysis \cite{vanessen.2012}. Subjects exhibiting excessive head motion were also removed. The preprocessed images were then parcellated using the Automated Anatomical Labeling (AAL) atlas into 116 non-overlapping anatomical regions, and voxel-wise signals within each region were averaged to obtain regional time series. Additional details of the imaging and preprocessing procedures can be found in our previous study \cite{huang.2020.NM}.

\subsection{Spatial Scaffold}  
\begin{figure}[t]
\centering

\begin{minipage}[t]{0.45\linewidth}
    \centering
    \includegraphics[width=\linewidth]{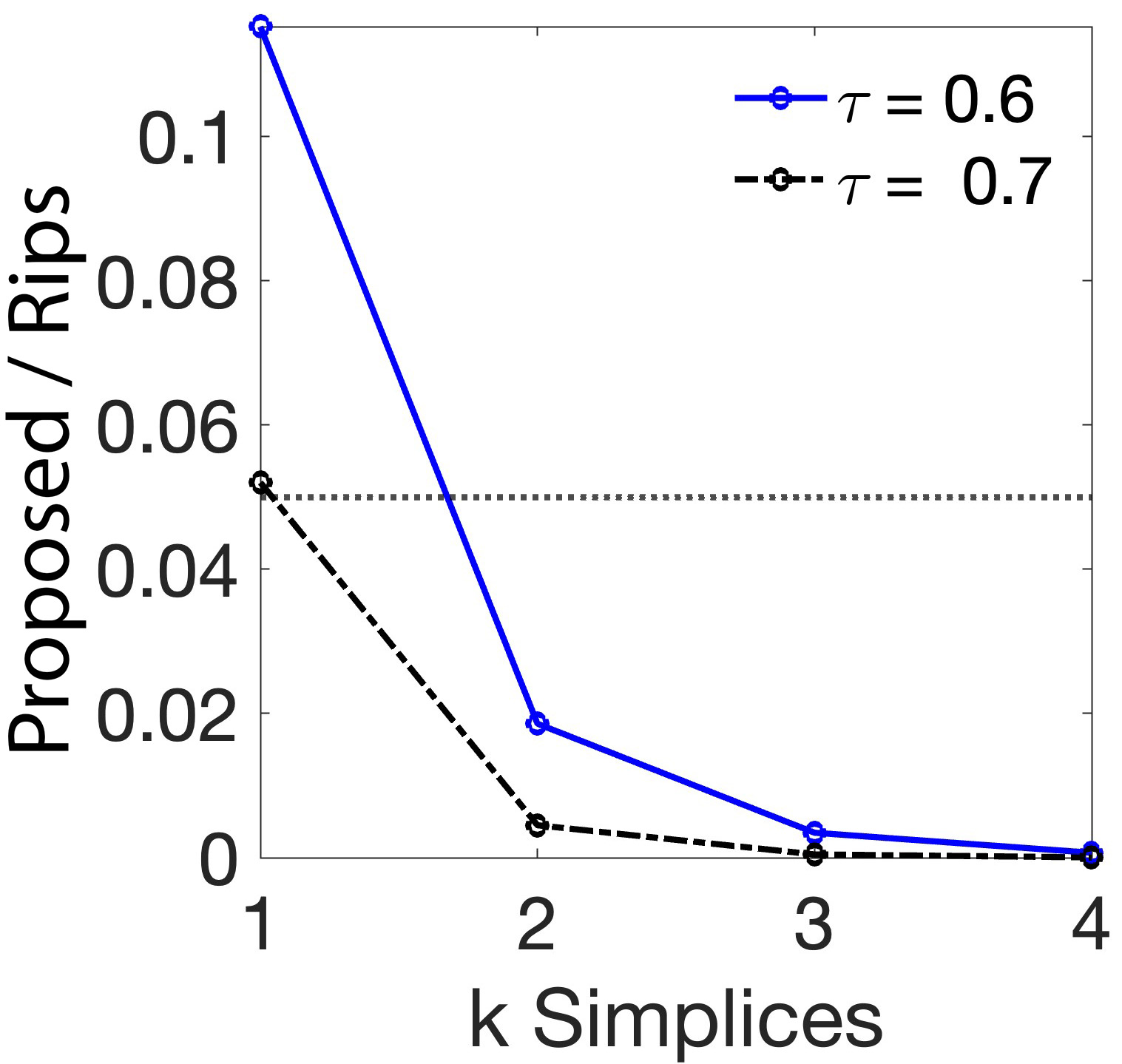}
\end{minipage}\hfill
\begin{minipage}[t]{0.53\linewidth}
    \centering
    \includegraphics[width=\linewidth]{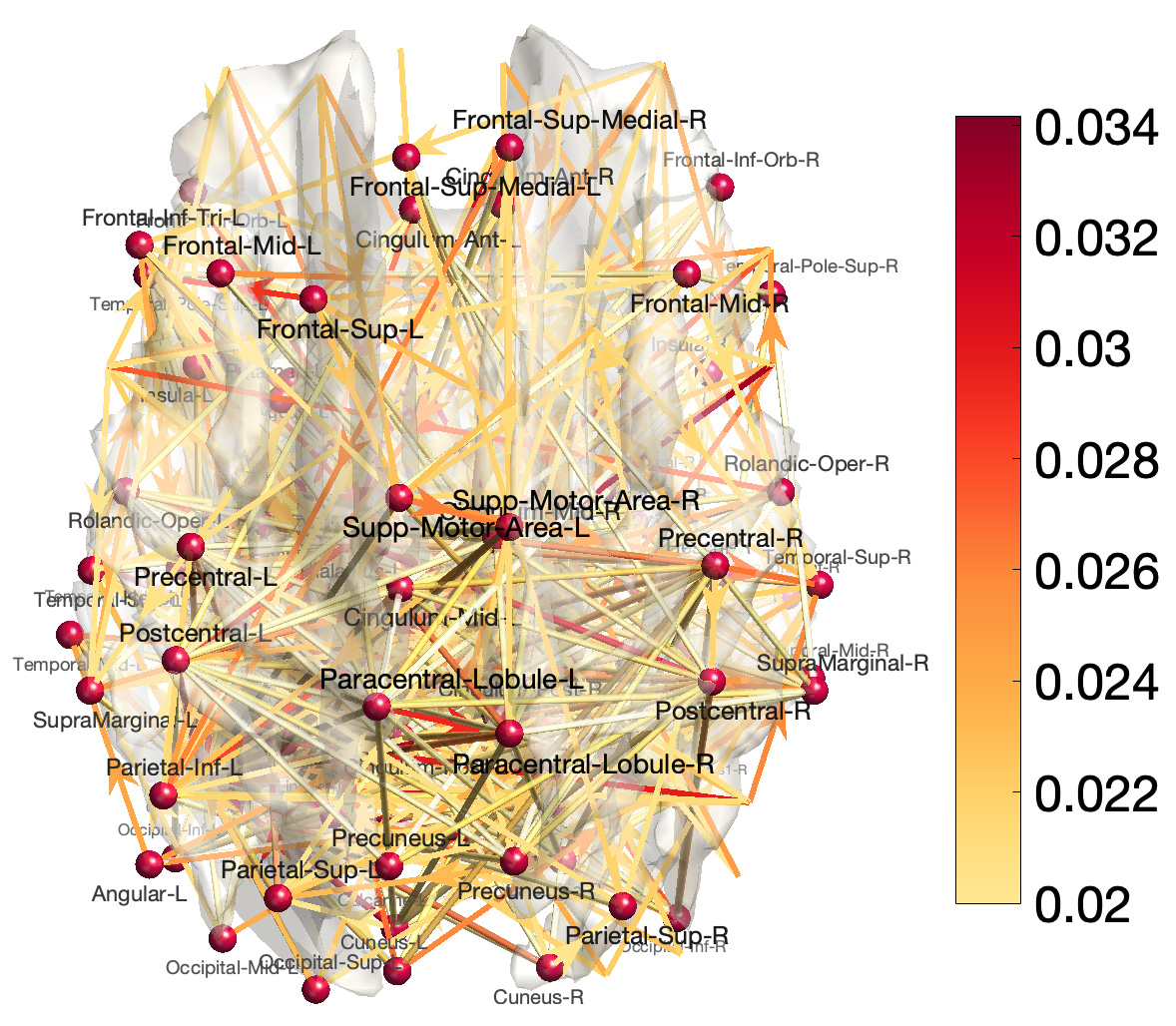}
\end{minipage}
\vspace{-0.25cm}
\caption{  
{\bf Left:} Average fraction of simplices in the Spatial Scaffold relative to the Rips complex (theoretical upper bound) across subjects in rs-fMRI connectivity.  
{\bf Right:} Average time-lagged correlation across all time points and subjects. The mean correlation is extremely low (maximum $r = 0.034$), indicating that none of the connections reach statistical significance.
}
\label{fig:averagecorrelation}
\vspace{-0.25cm}
\end{figure}

We constructed simplicial complexes from the AAL parcellation to capture not only pairwise but also higher-order motifs such as loops.  
AAL parcels served as {\it nodes}, with mean fMRI signals providing one representative time series per region.  Dynamic edge flows were estimated from resting-state fMRI using time-lagged Pearson correlations within 20-second sliding windows, corresponding to the minimum hemodynamic response scale \cite{keilholz.2017}.  This yielded a directed, time-varying matrix $X(t)=(X_{ij}(t))$, where $X_{ij}(t)$ denotes the influence of region $i$ on $j$.  Because of intersubject and temporal asynchrony, averaging $X(t)$ cancels directional effects, producing extremely weak mean connectivity (maximum $r=0.034$; Fig.~\ref{fig:averagecorrelation}, right).  To obtain a statistically robust directed flow, we used 2-simplices (triangles) to encode higher-order interactions.  Edges with weights below $\tau=0.6$ were removed, and directed triangles were included only when all three constituent edges exceeded $\tau$, ensuring hierarchical nesting and directional consistency across scales.

To manipulate simplicial complexes algebraically, we used {\it boundary (incidence) matrices}, which encode how simplices assemble across dimensions: $\mathbf{B}_1 \in \mathbb{R}^{|V|\times|E|}$ for node–edge and $\mathbf{B}_2 \in \mathbb{R}^{|E|\times|F|}$ for edge–face incidence \cite{edelsbrunner.2010,huang.2023}. 
These matrices provide a compact algebraic representation of higher-order interactions.  
Complexes and their corresponding boundary matrices were constructed hierarchically using the Intersecting Neighbor Sets algorithm \cite{schank.2005}, enabling scalable computation.

Unlike the commonly used Rips complex \cite{edelsbrunner.2010} in topological data analysis (TDA), which enumerates all possible $k$-simplices and rapidly becomes intractable due to combinatorial explosion \cite{choudhary.2021}, our approach constructs a sparse, data-driven complex that retains only statistically significant higher-order interactions.  The resulting Spatial Scaffold contains fewer than 0.02\% of the simplices generated by a full Rips complex (Fig.~\ref{fig:averagecorrelation}-left), providing a scalable and biologically meaningful topological representation.

\subsection{Dirichlet potential energy of network}  
We construct a causal framework based on the Dirichlet energy of time-varying edge flows $X(t)$.  
The 1-Hodge Laplacian \cite{anand.2023.TMI}. 
\[
\mathcal{L}_1 = \mathbf{B}_1^\top \mathbf{B}_1 + \mathbf{B}_2 \mathbf{B}_2^\top
\]
acts on edges, with $\mathbf{B}_1^\top \mathbf{B}_1$ capturing divergence (source–sink activity) and $\mathbf{B}_2 \mathbf{B}_2^\top$ capturing rotationality (cyclic circulation around triangles).  
For comparison, the traditional graph Laplacian, $\mathcal{L}_0 = \mathbf{B}_1 \mathbf{B}_1^\top$, actes on vertices and captured only acyclic pairwise interactions.  The associated Dirichlet potential energy is defined as
\[
E(X) = \tfrac{1}{2}\langle X, \mathcal{L}_1 X\rangle 
      = \tfrac{1}{2}\|\mathbf{B}_1 X\|_2^2 + \tfrac{1}{2}\|\mathbf{B}_2^\top X\|_2^2,
\]
which jointly measured divergence and cyclic circulation.  
This defines the energy landscape on which causal flows are modeled.

\begin{figure}[t]
\centering
\includegraphics[width=1\linewidth]{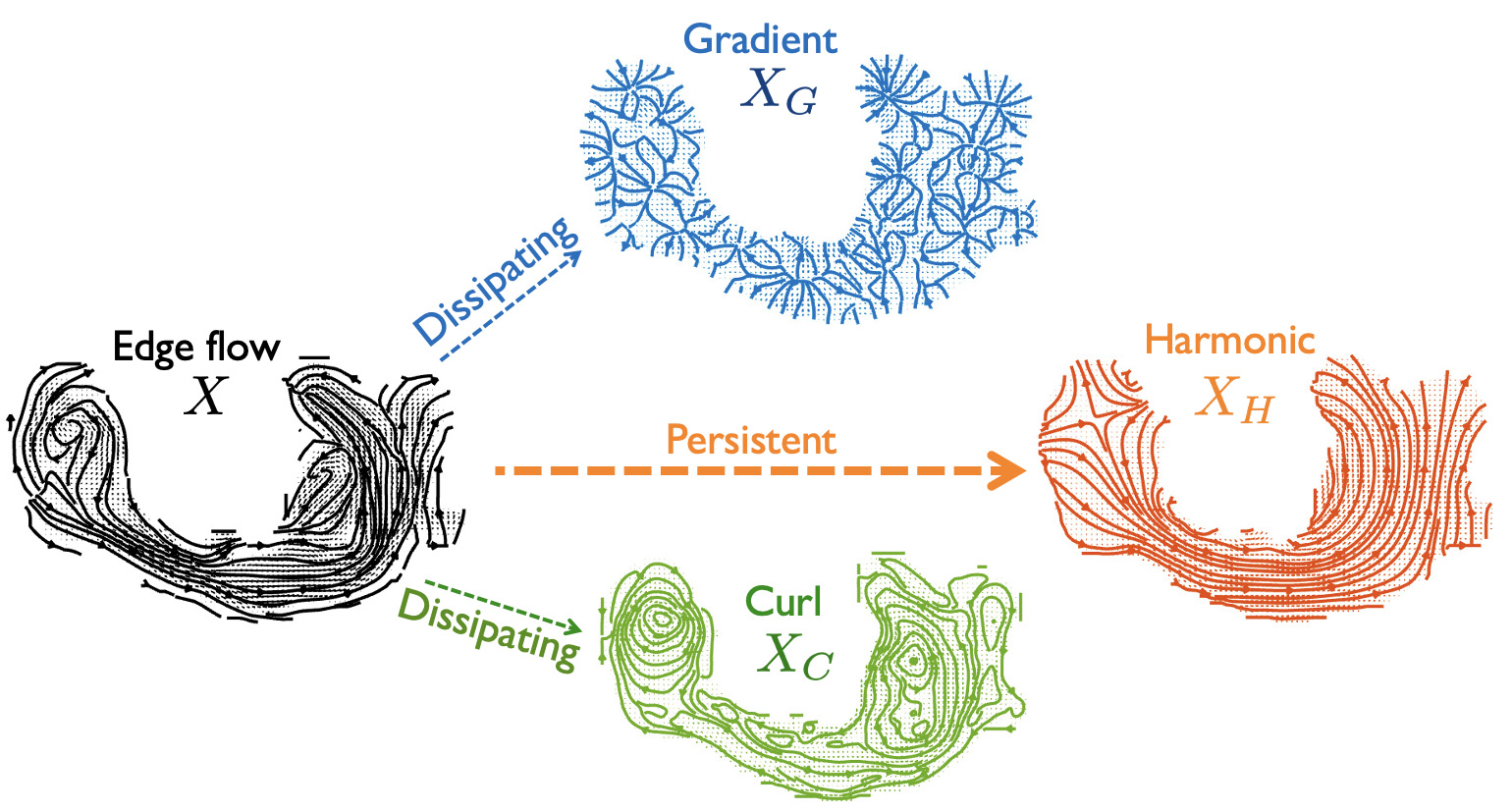}
\vspace{-0.75cm}
\caption{  Minimization of the Dirichlet potential energy of edge flow $X$ converges to the persistent harmonic flow $\red{X_H}$, which captures long-range, stable, and recurrent communication patterns.
The residual $X - \red{X_H}$, composed of the gradient $\blue{X_G}$ and curl $\green{X_C}$, represents transient dynamics that dissipate over time.}
\label{fig:HDC}
\end{figure}

An edge flow $X$ can be interpreted as a fluid-like transport of information, with causality corresponding to the system’s intrinsic tendency to evolve toward minimum-energy configurations (Fig. \ref{fig:HDC}). Minimization of $E(X)$ therefore provided a variational principle for causal propagation. The gradient flow of the potential energy yields Dirichlet diffusion through Onsager’s principle \cite{noirhomme.2024}:
\bqn
\frac{dX(t)}{dt} \;=\; -\,\nabla E(X(t)) \;=\; -\,\mathcal{L}_1 X(t).
\label{eq:diffusion}
\eqn
Along this trajectory, the energy decreases monotonically in steepest descent:
\[
\frac{d}{dt}E(X(t)) 
= \Big\langle \frac{dX}{dt}, \nabla E(X(t)) \Big\rangle 
= -\|\mathcal{L}_1 X(t)\|^2 \;\leq\; 0,
\]
ensuring that flows converges toward steady state $\red{X_H}$.  

Using the spectral decomposition of the 1-Hodge Laplacian \cite{su.2024}
$\mathcal{L}_1 \boldsymbol{\phi}_k = \lambda_k \boldsymbol{\phi}_k$ with 
$\lambda_k \ge 0$ \cite{anand.2023.TMI}, edge flow $X(t)$ admits solution
\[
X(t) = e^{-\mathcal{L}_1 t} X(0)
      = \sum_k e^{-\lambda_k t}\alpha_k \boldsymbol{\phi}_k, X(0) = \sum_k \alpha_k \boldsymbol{\phi}_k.
\qquad 
\]
As $t \to \infty$, all modes with $\lambda_k>0$ vanish while 
those with $\lambda_k=0$ remain:
\[
\red{X_H} = \mathcal{P}_H X(0),
\quad 
\mathcal{P}_H = \sum_{\lambda_k=0} 
\boldsymbol{\phi}_k \boldsymbol{\phi}_k^\top.
\]
$\red{X_H}$ thus captures the persistent, non-dissipative backbone of information 
flow—the portion of network dynamic that remains  stable after transient dynamics decay 
(Fig.~\ref{fig:HDC})\footnote{Matlab codes for computing harmonic flow is given in \url{https://github.com/laplcebeltrami/PH-STAT}.}.

\subsection{Counterfactual analysis on harmonic flow}  

Counterfactual analysis \cite{vanderweele.2020,dhinagar.2024,matsui.2022} provides a principled framework to probe how causal organization reconfigures under hypothetical interventions or disruptions that cannot be performed in vivo.  
Such interventions can be simulated at multiple scales: at the edge level by perturbing individual 
connections; at the motif level by damping or removing recurrent cycles to assess their stability; 
and at the system level by evaluating changes in the total Dirichlet energy.  

We formulate a counterfactual operator $\mathcal{C}$ to act linearly on the edge flow $X$, 
producing a modified flow $X^{(\mathcal{C})} = \mathcal{C}X$ that scales, deletes, or alters 
selected connections—for example, by setting an edge weight to zero, damping a cycle by 
a factor $0 \leq \alpha \leq 1$, or attenuating all edges incident to a hub node.  
Comparing $X^{(\mathcal{C})}$ with the baseline $X$ reveals how information flow reorganizes 
across the network and quantifies its resilience or restructing. 

The corresponding change in Dirichlet energy provides a global measure of network stability:
\[
\Delta E = E(X^{(\mathcal{C})}) - E(X)
         = \tfrac{1}{2}\big(X^{(\mathcal{C})\top}\mathcal{L}_1 X^{(\mathcal{C})} - X^\top \mathcal{L}_1 X\big).
\]
Large positive $\Delta E$ indicates increased dissipation and fragility, 
whereas small or near-zero $\Delta E$ implies that causal flow is rerouted along 
energetically equivalent pathways, reflecting robust topological compensation.

For harmonic flows $\red{X_H}$, which satisfy $\mathcal{L}_1 \red{X_H} = 0$, the Dirichlet energy 
is zero ($E(\red{X_H})=0$), representing non-dissipative, self-sustaining cycles of information flow.  
Counterfactual analysis of $\red{X_H}$ isolates the brain’s most stable communication backbone—
the part of causal organization that persists even when transient or dissipative components 
(gradient and curl flows) are removed.  
By perturbing $\red{X_H}$, we can test how these long-range recurrent pathways reorganize 
under hypothetical disruptions, providing a direct measure of the system’s intrinsic 
resilience and compensatory capacity.

When acted on by a counterfactual operator $\mathcal{C}$, the resulting flow
\[
\red{X_H}^{(\mathcal{C})} = \mathcal{C}(\mathcal{P}_H X) = \mathcal{P}_H(\mathcal{C}X)
\]
remains within the harmonic subspace and preserves its zero-energy property ($\Delta E = 0$).   
Hence, harmonic counterfactuals reorganize how information circulates 
without changing the global energy balance—yielding energetically conserved but 
topologically distinct configurations.  
These invariant harmonic flows form the persistent, self-correcting backbone of brain 
communication and provide a principled foundation for quantifying redundancy, 
compensation, and resilience in functional networks.

\section{Applications: Temporal Lobe Epilepsy}

We applied the proposed framework to model counterfactual disruptions and interventions in temporal lobe epilepsy (TLE) —a disorder characterized by recurrent synchronization and impaired large-scale stability \cite{chung.2023.NI}.  
Our analysis examined how disease-like perturbations reshape persistent causal pathways (harmonic flow, $\red{X_H}$) while preserving overall energy.

\begin{figure*}[t]
\centering
\includegraphics[width=0.8\linewidth]{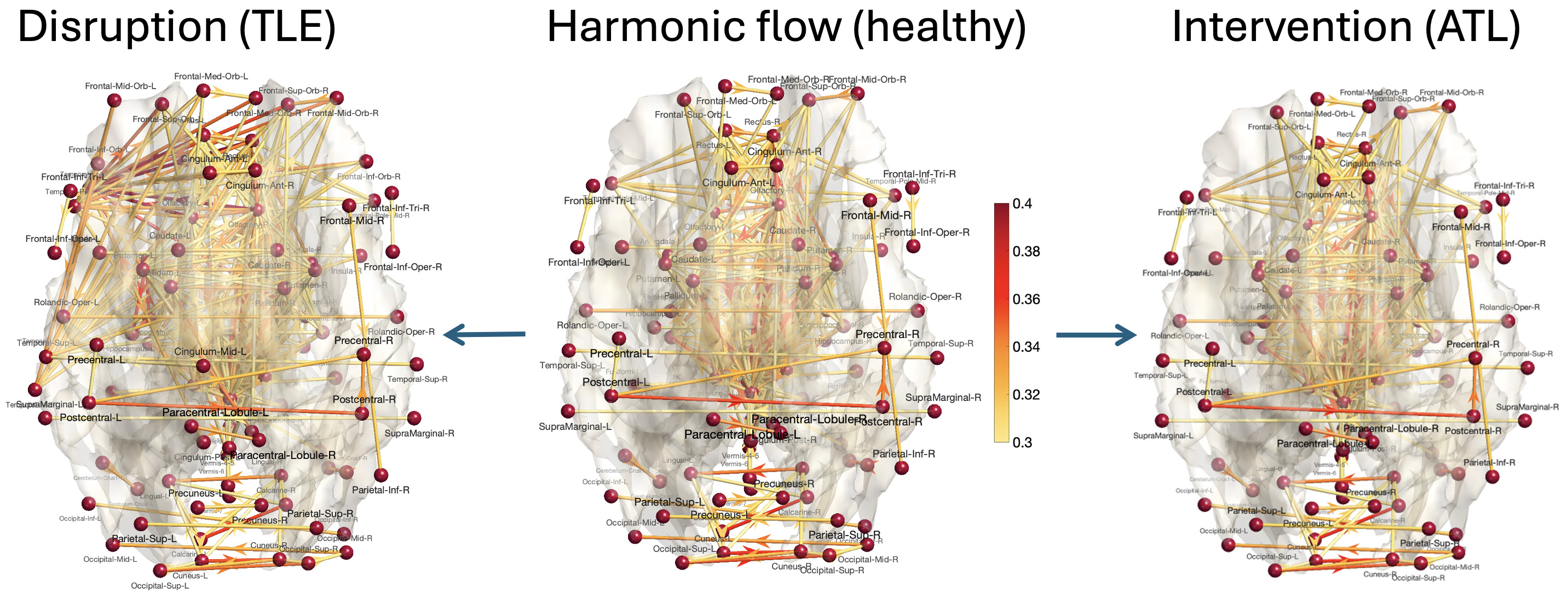}
\vspace{-0.25cm}
\caption{  
\textbf{Middle:} Ten dominant average harmonic flows ($\red{X_H}$) across time and 400 healthy subjects, showing strong recurrent interhemispheric and predominantly homotopic organization—most evident in visual (Calcarine, Cuneus, Lingual) and sensorimotor (Postcentral, Paracentral) regions.  
\textbf{Left:} After counterfactual perturbation of the temporal–limbic feedback loop, pathological hyperrecurrence shifts dominance from interhemispheric sensory pathways toward olfactory–limbic–vermal circuits.  
\textbf{Right:} After counterfactual simulation of left anterior temporal lobectomy (ATL) on healthy controls, the global topology remains largely preserved, indicating compensatory stabilization through contralateral and midline pathways.}
\label{fig:X_H}
\end{figure*}

\subsection{Pathological Disruption (Disease Model)}

TLE involves excessive synchronization within mesial temporal and limbic structures, including the hippocampus, amygdala, and orbitofrontal cortex \cite{bartolomei.2017}.  
We simulated this pathology using a harmonic counterfactual that amplified recurrent flows within these regions by 30\%, enhancing temporal$\to$ limbic drive consistent with patient evidence of hyper    connectivity \cite{gonzalez.2019}.  

Although the total Dirichlet energy remained unchanged ($\Delta E = 0$), the topology of persistent flows reorganized substantially (Table~\ref{tab:top10}, Fig.~\ref{fig:X_H}).  
In the healthy baseline, dominant harmonic flows were primarily interhemispheric sensory loops (Calcarine, Cuneus, Postcentral) and vermian–rectal connections.  
After perturbation, these were supplanted by stronger limbic and subcortical pathways, notably Olfactory$\to$Amygdala, Rectus$\to$Amygdala, and Amygdala$\to$Olfactory, reflecting hyperactive feedback within temporal–limbic hubs.  

This shift from symmetric sensory exchange to asymmetric limbic–subcortical recurrence mirrors electrophysiological and neuroimaging findings in TLE, where seizures propagate through hippocampal–amygdalar circuits and engage subcortical regions including the cerebellum and pallidum \cite{gonzalez.2019,jo.2019}.

\begin{table}[t!]
\centering
\caption{
Top 10 dominant harmonic flows before and after counterfactual amplification of the temporal–limbic feedback loop. 
Pathological recurrence strengthens limbic and subcortical interactions (e.g., Olf$\to$Amyg, Rect$\to$Amyg) while reducing the dominance of interhemispheric sensory loops. 
In contrast, simulated anterior temporal lobectomy (ATL) in healthy controls yields minimal reorganization, reflecting preserved global topology through contralateral and midline compensation.}
\label{tab:top10}
\setlength{\tabcolsep}{3pt}
\renewcommand{\arraystretch}{1.05}
\begin{tabular*}{\linewidth}{@{\extracolsep{\fill}} l c l c | l c l c @{}}
\hline
\multicolumn{4}{c|}{\textbf{Baseline (Healthy)}} & 
\multicolumn{4}{c}{\textbf{Disease Model (TLE)}} \\
\cline{1-8}
From & $\to$ & To & Mag. & From & $\to$ & To & Mag. \\
\hline
Vrm1–2 & $\to$ & Vrm10 & 0.38 & Olf-R & $\to$ & Amyg-R & 0.38 \\
Olf-R  & $\to$ & Vrm10 & 0.38 & Vrm1–2 & $\to$ & Vrm10 & 0.38 \\
Olf-R  & $\to$ & Vrm1–2 & 0.36 & Olf-R & $\to$ & Vrm10 & 0.38 \\
Calc-L & $\to$ & Calc-R & 0.36 & Rect-R & $\to$ & Amyg-R & 0.37 \\
Vrm10  & $\to$ & Rect-R & 0.36 & Amyg-L & $\to$ & Olf-R & 0.36 \\
Cune-L & $\to$ & Cune-R & 0.36 & Olf-R & $\to$ & Vrm1–2 & 0.36 \\
Olf-R  & $\to$ & Rect-R & 0.35 & Vrm10 & $\to$ & Rect-R & 0.36 \\
Post-L & $\to$ & Post-R & 0.35 & Calc-L & $\to$ & Calc-R & 0.36 \\
Rect-R & $\to$ & Vrm1–2 & 0.35 & Cune-L & $\to$ & Cune-R & 0.36 \\
Vrm10  & $\to$ & Pall-L & 0.35 & Olf-R & $\to$ & Rect-R & 0.35 \\
\hline
\end{tabular*}
{\it Abbreviations:} Olf = Olfactory, Amyg = Amygdala, Calc = Calcarine, 
Cune = Cuneus, Post = Postcentral, Rect = Rectus, Pall = Pallidum, Vrm = Vermis
\vspace{-0.25cm}
\end{table}

\subsection{Invasive Therapeutic Intervention}

We examined the effects of left \emph{anterior temporal lobectomy (ATL)}—a surgical treatment for drug-resistant temporal lobe epilepsy (TLE) \cite{kim.2021,sainburg.2024}.  ATL often removes the anterior temporal neocortex and mesial temporal structures, including the hippocampus and amygdala, thereby disrupting the recurrent temporal–limbic feedback loops that sustain seizure propagation.  Since such resection cannot be performed in healthy participants, we simulated its functional impact by attenuating left temporal–limbic and hippocampal–amygdalar flows by 70\%, approximating the disconnection produced by surgery.

In healthy controls,  applying the left-ATL counterfactual left this top 10 harmonic flows largely unchanged  (Table~\ref{tab:top10}, Fig.~\ref{fig:X_H}-right), indicating that persistent harmonic organization in the normative brain is supported by contralateral (right) limbic and midline vermis circuits that maintain energetic balance despite unilateral disruption.  This pattern reflects hemispheric redundancy and compensatory topology typical of healthy networks, where removal of one temporal pole minimally perturbs global harmonic equilibrium.  

By contrast, in the TLE disease model, the same perturbation induces a major redistribution toward limbic and subcortical dominance, demonstrating reduced compensatory capacity in pathological networks.

\section{Conclusion \& Discussion}

We proposed a unified framework for \emph{counterfactual causal analysis} that models pathological disruptions and therapeutic interventions as energy perturbations on network flows.  Within the Hodge framework, the harmonic flow $\red{X_H}$ emerged as the non-dissipative backbone of brain communication, allowing perturbations to reveal how stable recurrent pathways reorganize under disruption.  

Applied to temporal lobe epilepsy (TLE), pathological recurrence amplified temporal–limbic and subcortical feedback loops, whereas simulated anterior temporal lobectomy (ATL) in healthy controls produced minimal change—reflecting contralateral and midline compensation.  
Harmonic counterfactuals thus provide a principled tool to quantify network resilience and redundancy, offering a general framework for modeling virtual lesions, stimulation, and adaptive reorganization in brain networks.

\paragraph*{Acknowledgment.}
This study is funded by NIH MH133614 and NSF DMS-2010778. 


\end{document}